\documentclass[twocolumn,superscriptaddress,letter]{revtex4}%
\usepackage{amssymb}
\usepackage{color}
\usepackage{graphicx}
\usepackage{dcolumn}
\usepackage{bm}
\usepackage[header,title,page,titletoc]{appendix}
\usepackage{amsmath}
\usepackage{amsfonts}%
\providecommand{\U}[1]{\protect \rule{.1in}{.1in}}
\begin{document}
\title{Emergent Supersymmetric Many-Body Systems in Doped $\mathrm{Z}_{2}$
Topological Spin Liquid of the Toric-Code Model}
\author{Jing He}
\affiliation{Department of Physics, Beijing Normal University, Beijing, 100875 P. R. China }
\author{Jing Yu}
\affiliation{Department of Physics, Liaoning Shihua University, Fushun, 113001, P. R. China}
\author{Xing-Hai Zhang}
\affiliation{Department of Physics, Beijing Normal University, Beijing, 100875 P. R. China }
\author{Su-Peng Kou}
\thanks{Corresponding author}
\email{spkou@bnu.edu.cn}
\affiliation{Department of Physics, Beijing Normal University, Beijing, 100875 P. R. China }

\begin{abstract}
In this paper, we studied the doped $\mathrm{Z}_{2}$ topological spin liquid
of the toric-code model. We found that the doped holes become supersymmetric
particles. The ground state of the doped $\mathrm{Z}_{2}$ topological spin
liquid becomes new matters of quantum states - supersymmetric Bose-Einstein
condensation or supersymmetric superfluid. As a result, this system provides a
unique example of manipulatable supersymmetric many-body system.

\end{abstract}

\maketitle

Doped holes in a spin model had become an important issue since the discovery
of high Tc superconductivity in cuprates\cite{pwa}. As the microscopic model
of the high Tc superconductivity - the $t-J$ model has been intensively
studied for several decades. Motivated by the experimental facts in the
high-$T_{c}$ cuprates, the spin-charge separation idea \cite{KRS} was a very
basic concept by introducing spinless \textquotedblleft
holon\textquotedblright \ of charge $e$ and neutral spin-$1/2$
\textquotedblleft spinon\textquotedblright \ as the essential building blocks
of the restricted Hilbert space. However, for a long range antiferromagnetic
(AF) order, there is no true spin-charge separation at all. People pointed out
that a single doped hole in the AF order can be a charged spin bag\cite{bag},
or Shraiman-Siggia dipole\cite{ss}, or localized object\cite{string,1hole}
from different points of view. The true spin-charge separation occurs only in
the quantum disordered spin states (people call them quantum spin liquid
states). And the doped spin liquid is always a superconducting order with holon-condensation.

On the other hand, in the last decade, several exactly solvable spin models
with $\mathrm{Z}_{2}$ topological spin liquid were found, such as the
toric-code model \cite{k1}, the Wen-plaquette model \cite{wen5,wen} and the
Kitaev model on a honeycomb lattice \cite{k2}. It becomes an interesting issue
to study the properties of doped spin liquid by doping holes to these exactly
solvable spin models. In Ref.\cite{mei}, the fermi liquid nature of doped
$\mathrm{Z}_{2}$ topological spin liquid is obtained. People also studied the
effect of doped holes in the gapless phase of the Kitaev model in
Ref.\cite{e1,e,you} and pointed out that the topological superconducting state
can be its ground state.

In this paper, after studying the non-perturbative properties of doped holes,
we found that a doped hole in the $\mathrm{Z}_{2}$ topological spin liquid of
the toric-code model becomes a supersymmetric holon and a universal feature of
the doped $\mathrm{Z}_{2}$ topological spin liquid is the emergence of
\emph{supersymmetry}. We found that the ground state of the doped
$\mathrm{Z}_{2}$ topological spin liquid is \emph{supersymmetric Bose-Einstein
condensation} (SBEC) state, of which there exist the fermionic Goldstone mode
- Goldstino which is the partner of the Bosonic Goldstone mode. And we can
manipulate the supersymmetry by tuning the transverse external field to the
system. After breaking the supersymmetry by the transverse external field, the
ground state becomes a new matter of a quantum state - \emph{supersymmetric
superfluid} - Bose-Einstein condensation (BEC) state for the bosonic holons
and superconducting (SC) state for the fermionic holons. That is, the doped
$\mathrm{Z}_{2}$ topological spin liquid provides a unique example of
supersymmetric many-body system.

Our starting point is the so-called \emph{t-toric-code model} that is
described by the following Hamiltonian
\begin{align}
H &  =H_{\mathrm{t}}+H_{\mathrm{tc}}+H_{\mathrm{I}},\\
H_{\mathrm{t}} &  =-t\sum_{\langle ij\rangle,\sigma}\mathcal{P}_{s}c_{\sigma
i}^{\dagger}c_{\sigma j}\mathcal{P}_{s}+\mu \sum \limits_{i,\sigma}c_{i\sigma
}^{\dagger}c_{i\sigma}\nonumber \\
H_{\mathrm{tc}} &  =-A\sum \limits_{i\in \mathrm{even}}{Z_{i}}-B\sum
\limits_{i\in \mathrm{odd}}{X_{i},}\nonumber \\
H_{\mathrm{I}} &  =h^{x}\sum \limits_{i}s_{i}^{x}+h^{y}\sum \limits_{i}s_{i}%
^{y}\nonumber
\end{align}
where ${Z_{i}=}s_{i}^{z}s_{i+\hat{e}_{x}}^{z}s_{i+\hat{e}_{x}+\hat{e}_{y}}%
^{z}s_{i+\hat{e}_{y}}^{z},$ ${X_{i}=}s_{i}^{x}s_{i+\hat{e}_{x}}^{x}%
s_{i+\hat{e}_{x}+\hat{e}_{y}}^{x}s_{i+\hat{e}_{y}}^{x}$ with $A>0$, $B>0.$
$\sigma$ are the spin-indices representing spin-up $(\sigma=\uparrow)$ and
spin-down $(\sigma=\downarrow)$ for the electrons. $\mu$ is the chemical
potential. $\left \langle {i,j}\right \rangle $ denote two sites on the
nearest-neighbor link. $\mathcal{P}_{s}$ projects the Hilbert space onto the
subspace of one electron per-site. $s_{i}^{x,y,z}$ are the spin operators of
the electrons. For simplify, we consider the case of $A=B=g$ in this paper.
$H_{\mathrm{I}}$ is the external field terms with $h^{x}>0$, $h^{y}>0$.

The ground state of $H_{\mathrm{tc}}$ denoted by ${Z_{i}\equiv+1}$ and
${X_{i}\equiv+1}$ at each site is a $\mathrm{Z}_{2}$ topological state. For
the toric-code model, the elementary excitations are $\mathrm{Z}_{2}$ vortex
(${Z_{i}=-1}$ at even sub-plaquette) and $\mathrm{Z}_{2}$ charge (${X_{i}=-1}$
at odd sub-plaquette). Fermions are the bound states of a pair of
$\mathrm{Z}_{2}$ vortex and $\mathrm{Z}_{2}$ charge. When we add the
transverse external field, the total spin Hamiltonian $H_{\mathrm{tc}%
}+H_{\mathrm{I}}$ cannot be solved exactly. However, for small value of the
external field, the ground state is still $\mathrm{Z}_{2}$ topological order
and the topological properties don't change.

\begin{figure}[ptb]
\includegraphics[width=0.4\textwidth]{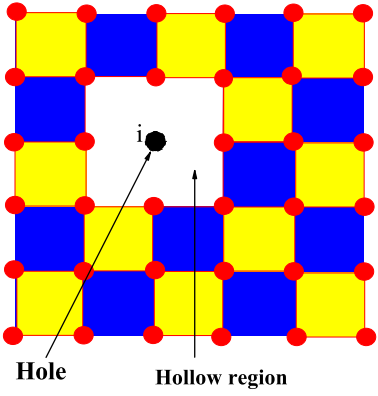}\caption{The illustration of a
single hole.}%
\end{figure}

\textit{Internal degree of freedom of a hole and emergent supersymmetry} : Now
we study a single hole doped in the toric-code model. As shown in Fig.1, a
hole is a charged vacancy to remove the spin degree of freedom at site $i,$ as
$c_{\sigma i}^{\dagger}\rightarrow c_{i}^{\dagger}$. Then near a hole, four
plaquette operators are forced to be zero, i.e. $Z_{i-\hat{e}_{x}}%
=Z_{i-\hat{e}_{y}}=X_{i}=X_{i-\hat{e}_{x}-\hat{e}_{y}}\equiv0.$ Around this
hole, there exist a boundary that separates the topological spin liquid
outside and a hollow area around the hole. For a hole in the $\mathrm{Z}_{2}$
topological order, the ground state degeneracy becomes $2$ (without being
considered the edge states)\cite{kou1,kou2}. To classify the degeneracy of the
ground states, as shown in Fig.2, we define two types of closed string
operators, $W_{v}(\mathcal{C}_{A})=\prod_{\mathcal{C}}s_{i}^{x}$ and
$W_{f}(\mathcal{C}_{B})=\prod_{\mathcal{C}}s_{i}^{y}.$ $W_{v}(\mathcal{C}%
_{A})$ is a closed string operator around a hole. $W_{f}(\mathcal{C}_{B})$ is
a closed string operator from the boundary of a hole to the boundary of the
system. Since $W_{v}(\mathcal{C}_{A})$ and $W_{f}(\mathcal{C}_{B})$ form the
Heisenberg algebra, we can map the Hilbert space of $W_{v}(\mathcal{C}_{A})$
and $W_{f}(\mathcal{C}_{B})$ onto that of a pseudo-spin ($S=\frac{1}{2}$)\ and
identify $W_{v}(\mathcal{C}_{A})$ and $W_{f}(\mathcal{C}_{B})$ to spin-1/2
operator $\mathcal{S}^{z}$ and $\mathcal{S}^{x}$ as $W_{v}(\mathcal{C}%
_{A})\rightarrow \mathcal{S}^{z}$ and $W_{f}(\mathcal{C}_{B})\rightarrow
\mathcal{S}^{x}$. Then, we use $\left \vert \uparrow \right \rangle $ and
$\left \vert \downarrow \right \rangle $ to denote the pseudo-spin degree of
freedom of the quantum states of a hole. These two quantum states can be
characterized by the fermion parity (or the flux quanta of the $\mathrm{Z}%
_{2}$ vortex) of the hole : $\left \vert \uparrow \right \rangle $ corresponds to
the state with even fermion parity, $\left \vert \downarrow \right \rangle $
corresponds to the state with odd fermion parity. Now the hole's statistics
dependents on its fermion parity.

For the zero external field case, $h^{x}=h^{y}=0$, the two quantum states
$\left \vert \uparrow \right \rangle $ and $\left \vert \downarrow \right \rangle $
are degenerate exactly. When we add the external field, the quantum tunneling
effect will lead to an energy splitting. Then we can derive the effective
model of a hole by the quantum tunneling theory as$\mathcal{\ H}_{\mathrm{h}%
}\simeq \tilde{h}^{x}\mathcal{S}^{x}+\tilde{h}^{z}\mathcal{S}^{z}$ where
$\tilde{h}^{x}=\frac{2(h^{y})^{L_{x}}}{(-8g)^{Lx-1}}$ and $\tilde{h}^{z}%
=\frac{24(h^{x})^{12}}{(-4g)^{11}}$. Here $L_{x}$ is the distance between the
boundary of the hole and the boundary of the system. See the detailed
calculations in Ref.\cite{kou1,yu}. In general, for an infinite system, we
have $\tilde{h}^{x}=0$. In addition, for the two-hole case, we have the
effective model of two holes as
\begin{equation}
\mathcal{H}_{\mathrm{h}}\simeq J^{xx}\mathcal{S}_{1}^{x}\mathcal{S}_{2}%
^{x}+J^{zz}\mathcal{S}_{1}^{z}\mathcal{S}_{2}^{z}+\sum \limits_{l=1,2}\tilde
{h}_{l}^{z}\mathcal{S}_{l}^{z}%
\end{equation}
where $J^{xx}=\frac{2(h^{y})^{l}}{(-8g)^{l-1}},$ $J^{zz}\simeq \frac
{(h^{x})^{L_{zz}}}{(-4g)^{L_{zz}-1}}$. $l$ is the distance between the holes
and $L_{zz}$ is the distance for a path surrounding the two holes. Due to
$J^{zz}\ll J^{xx},$ the effective model of the two-hole case is reduced into
\begin{equation}
\mathcal{H}_{\mathrm{h}}\simeq J^{xx}\mathcal{S}_{1}^{x}\mathcal{S}_{2}%
^{x}+\tilde{h}^{z}\mathcal{S}_{1}^{z}+\tilde{h}^{z}\mathcal{S}_{2}^{z}.
\end{equation}
Due to the term of $J^{xx}\mathcal{S}_{1}^{x}\mathcal{S}_{2}^{x},$ two holes
may exchange their fermion parities.

\begin{figure}[ptb]
\includegraphics[width=0.45\textwidth]{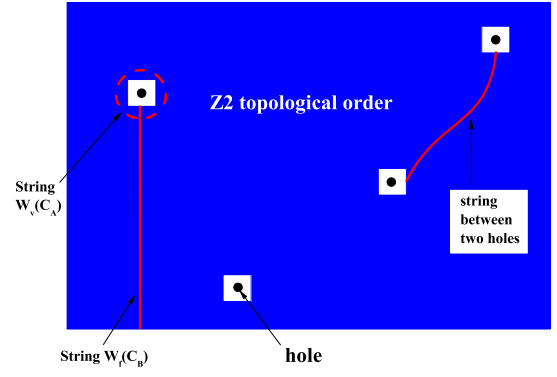}\caption{The illustration of
holes in a Z2 topological order of the toric-code model.}%
\end{figure}

So one may tune each parameter in above effective Hamiltonian $\mathcal{H}%
_{\mathrm{h}}$ by controlling the external field along special direction and
then manipulate the internal degree of freedom of holes. For example, in the
limit $J^{xx}\rightarrow0$, the quantum state with lowest energy is
$\mid \downarrow \rangle_{1}\otimes \mid \downarrow \rangle_{2}$ with odd fermion
parity for each hole. In the limit, $\tilde{h}^{z}\rightarrow0,$ the
eigenstates are $\left(  \mid \uparrow \rangle_{1}+\mid \downarrow \rangle
_{1}\right)  \otimes \left(  \mid \uparrow \rangle_{2}+\mid \downarrow \rangle
_{2}\right)  $ and $\left(  \mid \uparrow \rangle_{1}-\mid \downarrow \rangle
_{1}\right)  \otimes \left(  \mid \uparrow \rangle_{2}-\mid \downarrow \rangle
_{2}\right)  $. Now the average fermion parity of each hole is $1/2$ for both eigenstates.

\textit{Effective supersymmetric model:} Because the original statistics of a
hole is fermion, a holon (a hole in the $\mathrm{Z}_{2}$ topological state) is
a boson for the internal state $\left \vert \downarrow \right \rangle $ with odd
fermion parity or a fermion for the internal state $\left \vert \uparrow
\right \rangle $ with even fermion parity. Then for a single holon, we can use
a two-component supersymmetric operator to describe it as $c_{\tilde{\sigma}%
i}.$ $c_{\tilde{\sigma}i}$ plays the role of a two-component spinor with a
pseudo-spin where $c_{\tilde{\uparrow}i}=b_{i}$ and $c_{\tilde{\downarrow}%
i}=f_{i}$ are the charged boson and charged fermion operators on site $i.$

Now we consider the t-toric-code model with finite hole-concentration, of
which the effective Hamiltonian is $H_{\mathrm{eff}}=H_{\mathrm{t}%
}+\mathcal{H}_{\mathrm{h}}$ where $H_{\mathrm{t}}$ is the hopping term of
supersymmetric holons,
\begin{equation}
H_{\mathrm{t}}=-t\sum_{\langle ij\rangle \tilde{\sigma}}\mathcal{P}%
_{s}c_{\tilde{\sigma}i}^{\dagger}c_{\tilde{\sigma}j}\mathcal{P}_{s}+\mu
_{B}\sum_{i}b_{i}^{\dagger}b_{i}+\mu_{F}\sum_{i}f_{i}^{\dag}f_{i}.
\end{equation}
$\mathcal{P}_{s}$ projects the Hilbert space onto the subspace of one particle
per site. Now the projector $\mathcal{P}_{s}$ has no effect for the spinless
fermionic holon $f_{i}^{\dag}$ while guarantees the particle number of bosonic
holon $n_{i}^{b}=b_{i}^{\dag}b_{i}$. $\mu_{B}$ and $\mu_{F}$ are the chemical
potentials for the bosonic holons and that of fermionic holon, respectively.

To characterize the supersymmetry, we introduce the generators $\mathcal{S}%
_{i}^{x},$ $\mathcal{S}_{i}^{y}$, $\mathcal{S}_{l}^{z}$ that are just the
string operators $W_{f}(\mathcal{C}_{B}),$ $-iW_{f}(\mathcal{C}_{B}%
)W_{v}(\mathcal{C}_{A}),$ $W_{v}(\mathcal{C}_{A})$. And we have $\mathcal{S}%
_{l}^{x}=(\mathcal{S}_{i}^{-}+\mathcal{S}_{i}^{+})/2$, $\mathcal{S}_{l}%
^{y}=(\mathcal{S}_{i}^{-}-\mathcal{S}_{i}^{+})/2i.$ It is clear that
$\mathcal{S}_{i}^{\pm}$ is a fermionic operator. Physically, $\mathcal{S}%
_{i}^{-}$ turns a fermionic holon into a bosonic holon, and $\mathcal{S}%
_{i}^{+}$ does the opposite. By the two-component supersymmetric operator, we
have $\mathcal{S}^{-}$ $=b_{i}^{\dag}f_{i}$, $\mathcal{S}_{i}^{+}=b_{i}%
f_{i}^{\dag}$ and $\mathcal{S}_{l}^{z}=(b_{i}^{\dagger}b_{i}-f_{i}^{\dag}%
f_{i})/2$. Thus we re-write the term $\mathcal{H}_{\mathrm{h}}$ into
\begin{equation}
\mathcal{H}_{\mathrm{h}}\simeq \sum_{ij}J_{l}^{xx}\mathcal{S}_{i}%
^{x}\mathcal{S}_{j}^{x}+\frac{\tilde{h}^{z}}{2}\sum_{i}(b_{i}^{\dagger}%
b_{i}-f_{i}^{\dag}f_{i})
\end{equation}
where $J_{l}^{xx}=\frac{2(h^{y})^{l}}{(-8g)^{l-1}}$. Because the exchange term
$J_{l}^{xx}$ decays exponentially, we may only consider the shortest case that
is $l=2$ (for $l<2,$ the hollow regions of two holons merge and we cannot
define the corresponding quantum tunneling process). As a result, the fermion
parity exchange term $\sum_{ij}J_{l}^{xx}\mathcal{S}_{i}^{x}\mathcal{S}%
_{j}^{x}$ is reduced into
\begin{align}
&  J_{l=2}^{xx}\sum_{i}(\mathcal{S}_{i}^{x}\mathcal{S}_{i+3e_{x}}%
^{x}+\mathcal{S}_{i}^{x}\mathcal{S}_{i+3e_{y}}^{x})\\
&  =\frac{J_{l=2}^{xx}}{4}\sum_{i}[(b_{i}^{\dag}f_{i}+b_{i}f_{i}^{\dag
})(b_{i+3e_{x}}^{\dag}f_{i+3e_{x}}+b_{i+3e_{x}}f_{i+3e_{x}}^{\dag})\nonumber \\
&  +(b_{i}^{\dag}f_{i}+b_{i}f_{i}^{\dag})(b_{i+3e_{y}}^{\dag}f_{i+3e_{y}%
}+b_{i+3e_{y}}f_{i+3e_{y}}^{\dag})].\nonumber
\end{align}

Thus we have an exact supersymmetry for the toric code model without the
external field, i.e., $[\mathcal{S}_{i}^{-},$ $H_{\mathrm{eff}}]=[\mathcal{S}%
_{i}^{+},$ $H_{\mathrm{eff}}]=0.$ When there exists the external field
($h^{x}\neq0,$ $h^{y}\neq0$), the supersymmetry is broken explicitly, i.e.,
$[\mathcal{S}_{i}^{-},$ $H_{\mathrm{eff}}]\neq0,$ $[\mathcal{S}_{i}^{+},$
$H_{\mathrm{eff}}]\neq0$.

\textit{Supersymmetric Bose-Einstein condensation}: We firstly study the
many-body system with exact supersymmetry as $[\mathcal{S}^{\pm},$ $H]=0.$ Now
for the dilute gas limit of holons (the hole-concentration $\delta$ is smaller
than $1\%$), the effective model is
\begin{equation}
H_{\mathrm{eff}}=-t\sum_{\langle ij\rangle}b_{i}^{\dagger}b_{j}-t\sum_{\langle
ij\rangle}f_{i}^{\dagger}f_{j}+\mu_{B}\sum_{i}b_{i}^{\dagger}b_{i}+\mu_{F}%
\sum_{i}f_{i}^{\dag}f_{i}%
\end{equation}
where $\mu_{F}=\mu_{B}=-4t$. $b_{j}$ denotes the annihilation operator of the
hard-core bosons. Because in the small hole-concentration limit, we can
release the single-occupied condition of the bosonic holons and consider the
holons as non-interacting bosons. For non-interacting bosons we always have
$\mu_{B}=-4t$ (we measure the energy from the bottom of single particle
dispersion) at zero temperature, regardless of boson number. Supersymmetry of
$H_{\mathrm{eff}}$ requires $\mu_{F}=\mu_{B}$, as a result we have $0$ or $1$
fermionic holon. Now each holon becomes a boson and the ground state is a BEC
state with holon-condensation as $\langle0|b_{\mathbf{k}=0}|0\rangle
=b_{0}e^{i\varphi_{0}}\neq0.$ Here $|0\rangle$ denotes the ground state. Thus
the ground state spontaneously$\ $breaks both global $U(1)$ symmetry and supersymmetry.

There are two types of collective excitations. One type is the Goldstone mode
that describes the phase fluctuations as $\varphi_{0}\rightarrow \varphi
_{0}+\varphi_{i}$. The effective Hamiltonian of the Goldstone mode is
$H=-2tb_{0}^{2}\sum_{\langle ij\rangle}\cos(\varphi_{i}-\varphi_{j})\ $which
is really a two-dimensional XY model. Another type of collective excitation is
the Goldstino mode which can be regarded as the fermionic "spin
wave"\cite{yue}. Due to the holon-condensation, we have $\mathcal{S}_{i}%
^{+}=b_{0}f_{i}^{\dag}.$ From the hopping term of $f_{i}^{\dag}$, we have the
effective model of the Goldstino as $H_{\mathrm{Goldstino}}=-t\left \vert
b_{0}\right \vert ^{-2}\sum_{\left \langle ij\right \rangle }\mathcal{S}_{i}%
^{+}\mathcal{S}_{j}^{-}.$ The dispersion of the Goldstino mode is given by
$E_{\mathbf{k}}=\left \vert b_{0}\right \vert ^{-2}\varepsilon_{\mathbf{k}}$
where $\varepsilon_{\mathbf{k}}=-2t(\cos k_{x}+\cos k_{y})$. Due to the
existence of the Goldstino mode, we call the unique ground state to be
supersymmetric Bose-Einstein condensation (SBEC).

\textit{Supersymmetric Bose-Einstein condensation with Fermi surface}: For the
case of $h^{x}\neq0,$ $h^{y}=0,$ the effective model turns into
\begin{align}
H_{\mathrm{eff}}  &  =-t\sum_{\langle ij\rangle}b_{i}^{\dagger}b_{j}%
-t\sum_{\langle ij\rangle}f_{i}^{\dagger}f_{j}+\sum_{i}\mu_{B}b_{i}^{\dagger
}b_{i}\nonumber \\
&  +\sum_{i}\mu_{F}f_{i}^{\dag}f_{i}+\frac{\tilde{h}^{z}}{2}\sum_{i}%
(b_{i}^{\dagger}b_{i}-f_{i}^{\dag}f_{i})
\end{align}
where $\tilde{h}^{z}=\frac{24(h^{x})^{12}}{(-4g)^{11}}$. The supersymmetry is
partially broken by $\tilde{h}^{z}$ term.

Because there is no couple between fermionic holons and bosonic holons, the
ground state can be considered to be a Bose-Fermi mixture. The bosonic holons
condense and fermionic holons form Fermi liquid. In the dilute hole limit, we
can simplify the Fermi surface of the fermionic holons to be a circle with a
radius $k_{F}$. Now in continuum limit, the effective Hamiltonian can be
reduced into
\begin{equation}
H_{\mathrm{eff}}\simeq \sum_{\mathbf{k}}[\frac{\mathbf{k}^{2}}{2m}+\left(
\mu_{B}\right)  _{\mathrm{eff}}]b_{\mathbf{k}}^{\dagger}b_{\mathbf{k}}%
+\sum_{\mathbf{k}}[\frac{\mathbf{k}^{2}}{2m}+\left(  \mu_{F}\right)
_{\mathrm{eff}}]f_{\mathbf{k}}^{\dagger}f_{\mathbf{k}}%
\end{equation}
where the effective chemical potential of bosonic holons $\left(  \mu
_{B}\right)  _{\mathrm{eff}}$ and the effective chemical potential of
fermionic holons $\left(  \mu_{F}\right)  _{\mathrm{eff}}$ are $\left(
\mu_{B}\right)  _{\mathrm{eff}}=\mu_{B}+\frac{\tilde{h}^{z}}{2}$ and $\left(
\mu_{F}\right)  _{\mathrm{eff}}=\mu_{F}-\frac{\tilde{h}^{z}}{2},$
respectively. $m$ is the effective mass of holons, $m\simeq \frac{1}{2t}.$

Then we can estimate the number of fermionic holons and that of bosonic holons
by minimizing the total ground state energy. We define the total holon number
to be $N_{t}=N_{b}+N_{f}$ where $N_{b}=\sum_{i}n_{i}^{b}$ and $N_{f}=\sum
_{i}n_{i}^{f}.$ The total ground state energy is $E_{\mathrm{total}}%
=E_{b}+E_{f},$ where $E_{b}=\frac{\tilde{h}^{z}}{2}N_{b}=\frac{\tilde{h}^{z}%
}{2}(N_{t}-N_{f})$ is the ground state energy of bosonic holons and
$E_{f}=\left(  \delta_{f}\right)  ^{2}\frac{\pi N}{m}-\frac{N_{f}\tilde{h}%
^{z}}{2}$ is the ground state energy of fermionic holons. Here $N$ is the
lattice number and $\delta_{f}$ is the concentration of fermionic holons. From
the condition, $\frac{\partial E_{\mathrm{total}}}{\partial N_{f}}=0,$ we
obtain the concentration of fermionic holons as $\delta_{f}=\frac{m\tilde
{h}^{z}}{2\pi}=\frac{\tilde{h}^{z}}{4\pi t}.$ Because the maximum fermionic
holon's concentration is $\delta$ which is the hole concentration, we get a
critical value of the external field, $(\tilde{h}^{z})_{c}=4\pi t\delta$ or
$\left(  h^{x}\right)  _{c}=\left(  4g\right)  ^{11/12}\left(  \frac{\pi
t\delta}{6}\right)  ^{1/12}$.

\begin{figure}[ptb]
\includegraphics[width=0.5\textwidth]{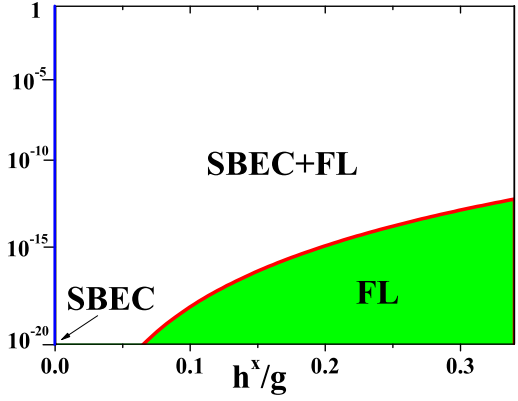}\caption{The phase diagram: the
supersymmetric Bose-Einstein-condensation (SBEC) (the blue line at $h^{x}=0$),
the supersymmetric Bose-Einstein-condensation with Fermi surface (SBEC+FL),
the Fermi liquid (FL). There is a quantum phase transition at $(h^{x})_{c}$
(the red line).}%
\label{Fig.4}%
\end{figure}

For small external field case, $h^{x}<(h^{x})_{c},$ there exists bosonic
holons with the concentration to be $\delta_{b}=\delta-\delta_{f}.$ Now the
ground state is a mixture of SBEC of bosonic holons and Fermi liquid (FL) of
fermionic holons (we call it SBEC+FL state). In general, $\tilde{h}^{z}$ is
very tiny. For example, for the case of $h^{x}/g=0.3,$ we have $\tilde{h}%
^{z}/g\simeq-3\times10^{-12}$. So we always have the SBEC state for bosonic
holons and the Fermi liquid for fermionic holons with very tiny fermi surface.

For large external field case, $h^{x}>(h^{x})_{c},$ all holons are fermions.
The ground state is a Fermi liquid (FL) state with the number of fermionic
holons as $N_{f}=N_{t}$. Now the effective chemical potential of fermionic
holons is $\left(  \mu_{F}\right)  _{\mathrm{eff}}\simeq2\pi t\delta$ and the
effective chemical potential of bosonic holons is $\left(  \mu_{B}\right)
_{\mathrm{eff}}\simeq \tilde{h}^{z}/2$. There exists a finite energy gap to
excite a bosonic holon. While the fermions have no energy gap.

From these results we plot the phase diagram in Fig.3, of which the red line
denotes the quantum phase transition between SBEC+FL and FL. The FL state is
only stable in the limit of $t\delta \rightarrow0$ and $h^{x}/g\rightarrow1$.
However, for $h^{x}>0.34g,$ a topological quantum phase transition occurs and
the ground state turns into a spin polarized state without topological
order\cite{pt1,pt2,pt5,pt3,pt4}. Thus we can only discuss the case of small
external field as $h^{x}<0.34g$.

\textit{Supersymmetric superfluid}: In this section we study the case of
$h^{y}\neq0,$ $h^{x}\neq0.$ The supersymmetry is broken completely. Now the
effective model is
\begin{align}
H  &  =-t\sum_{\langle ij\rangle}b_{i}^{\dagger}b_{j}-t\sum_{\langle
ij\rangle}f_{i}^{\dagger}f_{j}+\mu_{B}\sum_{i}b_{i}^{\dagger}b_{i}\nonumber \\
&  +\mu_{F}\sum_{i}f_{i}^{\dag}f_{i}+\frac{\tilde{h}_{l}^{z}}{2}\sum_{i}%
(b_{i}^{\dagger}b_{i}-f_{i}^{\dag}f_{i})\nonumber \\
&  +\frac{J_{l=2}^{xx}}{4}\sum_{i}[(b_{i}^{\dag}f_{i}+b_{i}f_{i}^{\dag
})(b_{i+3e_{x}}^{\dag}f_{i+3e_{x}}+b_{i+3e_{x}}f_{i+3e_{x}}^{\dag})\nonumber \\
&  +(b_{i}^{\dag}f_{i}+b_{i}f_{i}^{\dag})(b_{i+3e_{y}}^{\dag}f_{i+3e_{y}%
}+b_{i+3e_{y}}f_{i+3e_{y}}^{\dag})].
\end{align}
The ground state is really a BEC of bosonic holons $\langle0|b_{0}%
|0\rangle=b_{0}e^{i\varphi_{0}}$ with induced pairing of fermionic holons.

Let us estimate the induced SC pairing of fermionic holons. The effective
Hamiltonian of fermionic holons is
\begin{align}
H_{f} &  \simeq-t\sum_{\left \langle i,j\right \rangle }f_{i}^{\dag}%
f_{j}+\left(  \mu_{F}\right)  _{\mathrm{eff}}\sum_{i}f_{i}^{\dag}%
f_{i}\nonumber \\
&  +\frac{b_{0}^{2}J_{l=2}^{xx}}{4}\sum_{i}(f_{i}^{\dag}f_{i+3e_{x}}^{\dag
}+f_{i}^{\dag}f_{i+3e_{y}}^{\dag})+h.c.
\end{align}
Thus, when the bosonic holons condensate, there exists p$_{x}+$ p$_{y}$
superconducting order parameter for the non-interacting fermionic holon
\begin{equation}
\langle0|{f_{\mathbf{k}}^{\dagger}f_{-\mathbf{k}}^{\dagger}}|0\rangle
=\frac{b_{0}^{2}J_{l=2}^{xx}}{4}[(\sin3k_{x}+\sin3k_{y})].
\end{equation}
Now we have a gapless fermionic holon with p$_{x}+$ p$_{y}$ pairing.

Due to the condensation of the bosonic holons and the induced SC pairing of
fermionic holons, the effective model of the Goldstino mode turns into
\begin{align}
H_{\mathrm{Goldstino}} &  =-t\left \vert b_{0}\right \vert ^{-2}\sum
_{\left \langle ij\right \rangle }\mathcal{S}_{i}^{+}\mathcal{S}_{j}^{-}+\left(
\mu_{F}\right)  _{\mathrm{eff}}\left \vert b_{0}\right \vert ^{-2}\sum
_{i}\mathcal{S}_{i}^{+}\mathcal{S}_{i}^{-}\nonumber \\
&  +J_{l=2}^{xx}\sum_{i}(\mathcal{S}_{i}^{+}\mathcal{S}_{i+3e_{x}}%
^{+}+\mathcal{S}_{i}^{+}\mathcal{S}_{i+3e_{y}}^{+})+h.c..
\end{align}
Then the dispersion of the Goldstino mode is derive as $E_{\mathbf{k}%
}=\left \vert b_{0}\right \vert ^{-2}\sqrt{\varepsilon_{k}^{2}+\Delta_{k}^{2}}$
where $\varepsilon_{k}=-2t(\cos k_{x}+\cos k_{y})+\left(  \mu_{F}\right)
_{\mathrm{eff}}$ and $\Delta_{k}^{2}=(\frac{J_{l=2}^{xx}b_{0}^{2}}{4}%
)^{2}[(\sin3k_{x}+\sin3k_{y})^{2}].$

\textit{Conclusion} : In this paper, we found that for the $\mathrm{Z}_{2}$
topological spin liquid of the toric-code model, a doped hole becomes a
charged supersymmetric particle. Thus the doped $\mathrm{Z}_{2}$ topological
spin liquid of the toric-code model provides a unique example of
supersymmetric many-body system. The ground state is a new matter of quantum
state - SBEC, of which there exists fermionic Goldstone mode - Goldstino which
is the partner of Bosonic Goldstone mode. And we can tune the supersymmetry by
adding transverse external field to the toric-code model. After breaking the
supersymmetry by transverse external field, the ground state may be a
supersymmetric superfluid - BEC state for bosonic holons and SC state for
fermionic holons.

This work is supported by National Basic Research Program of China (973
Program) under the grant No. 2011CB921803, 2012CB921704 and NFSC Grant No. 11174035.


\begin{thebibliography}{99}                                                                                               %


\bibitem {pwa}P. W. Anderson, Science \textbf{235}, 1196 (1987).

\bibitem {KRS}S.A. Kivelson, D.S. Rokhsar, and J.R. Sethna, Phys. Rev.
\textbf{B 35}, 8865(1987).

\bibitem {bag}J. R. Schrieffer, X.-G. Wen, and S.-C. Zhang, Phys. Rev. Lett.
\textbf{60}. 944 (1988).

\bibitem {ss}B. I. Shraiman and E. D. Siggia, Phys. Rev. Lett. \textbf{62},
1564 (1989); B. I. Shraiman and E. D. Siggia, Phys. Rev. Lett. \textbf{61},
467 (1988).

\bibitem {string}D. N. Sheng, Y. C. Chen, and Z. Y. Weng, Phys. Rev. Lett.
\textbf{77}, 5102 (1996).

\bibitem {1hole}Z.Y. Weng, V.N. Muthukumar, D.N. Sheng, and C.S. Ting, Phys.
Rev. \textbf{B 63}, 075102 (2001).

\bibitem {k1}A. Kitaev, Ann. Phys. \textbf{303}, 2 (2003).

\bibitem {wen5}X. G. Wen, {Phys. Rev. Lett. }\textbf{90}, 016803 (2003).

\bibitem {wen}{X. G. Wen}, \emph{Quantum Field Theory of Many-Body Systems},
(Oxford Univ. Press, Oxford, 2004).

\bibitem {k2}A. Kitaev, Ann. Phys. \textbf{321}, 2 (2006).

\bibitem {mei}Jia-Wei Mei, Phys. Rev. Lett. \textbf{108}, 227207 (2012).

\bibitem {e1}A. J. Willans, J. T. Chalker, and R. Moessner, Phys. Rev. Lett.
\textbf{104}, 237203 (2010).

\bibitem {e}T. Hyart, A. R. Wright, G. Khaliullin, and B. Rosenow, Phys. Rev.
\textbf{B 85}, 140510(R) (2012).

\bibitem {you}Y.-Z. You, I. Kimchi, and A. Vishwanath, Phys. Rev. \textbf{B
86}, 085145 (2012).

\bibitem {kou1}S. P. Kou, Phys. Rev. Lett. \textbf{102}, 120402 (2009); Phys.
Rev. \textbf{A 80}, 052317 (2009).

\bibitem {yu}J. Yu and S. P. Kou, Phys. Rev. \textbf{B 80}, 075107 (2009).

\bibitem {kou2}S. P. Kou, M. Levin, and X. G. Wen, Phys. Rev. B \textbf{78},
155134 (2008).

\bibitem {yue}Yue Yu, Kun Yang, Phys. Rev. Lett. \textbf{100}, 090404 (2008);
Yue Yu and Kun Yang, Phys. Rev. Lett. \textbf{105}, 150605 (2010).

\bibitem {pt1}J. Vidal, S. Dusuel, K. P. Schmidt, Phys. Rev. \textbf{B 79},
033109 (2009).

\bibitem {pt2}I. S. Tupitsyn, A. Kitaev, N. V. Prokof'ev, and P. C. E. Stamp,
Phys. Rev. \textbf{B 82}, 085114 (2010).

\bibitem {pt5}S. Dusuel, M. Kamfor, R. Orus, K. P. Schmidt, J. Vidal, Phys.
Rev. Lett. \textbf{106}, 107203 (2011).

\bibitem {pt3}F. C. Wu, Youjin Deng, N. Prokof'ev, Phys. Rev. \textbf{B 85},
195104 (2012).

\bibitem {pt4}M. D. Schulz, S. Dusuel, R. Orus, J. Vidal, K. P. Schmidt, New
J. Phys. \textbf{14}, 025005 (2012).
\end{thebibliography}
\end{document}